\input{aipcheck}

\documentclass[
    ,final            
  ]
  {aipproc}

\layoutstyle{6x9}

\begin{document}
\newcommand{\mnras}{MNRAS}
\newcommand{\apj}{ApJ}
\newcommand{\aj}{AJ}
\newcommand{\apjl}{ApJL}
\newcommand{\apjs}{ApJS}
\newcommand{\aap}{A\&A}
\newcommand{\aaps}{A\&AS}
\newcommand{\rsol}{R$_\odot$}

\title{Spectropolarimetry of single and binary stars}

\author{Tim J Harries}{
  address={School of Physics, University of Exeter, Stocker Road,
  Exeter EX4 4QL.}
}



\begin{abstract}
  
  Spectropolarimetry is a photon-hungry technique that will reach
  fruition in the 8-m telescope age. Here I summarize some of the
  stellar spectropolarimetric research that my collaborators and I
  have undertaken, with particular emphasis on the circumstellar
  environment of massive stars, symbiotic binaries, and star
  formation.

\end{abstract}

\maketitle


\section{Introduction}

Stellar spectropolarimetry\footnote{Note that this paper
  focuses on the use of linear spectropolarimetry, although of course
  circular spectropolarimetry is an extremely powerful tool for
  probing stellar magnetic field structures, both at the stellar
  surface and in circumstellar material (e.g.
  \cite{2001MNRAS.326.1265D,2002MNRAS.333..339J}).} is a relatively
underused technique that has enormous diagnostic potential,
particularly for research into circumstellar matter and close binary
systems. The principle difficulty with the method is that very high
signal-to-noise observations are required, since typical astrophysical
polarimetric signatures may be on the order of a few tenths of a
percent. When photon noise is dominant, the polarimetric precision
$\Delta P$ goes as
\begin{equation}
\Delta P \approx  \frac{100}{\sqrt{N}}
\end{equation}
where $\Delta P$ is in percent and $N$ is the number of photons.  One
can see that a S/N of 1000 or more per pixel is needed to measure a
spectropolarimetric signal at the 0.1\% level. At moderate resolutions
($R \sim 5000$) this limits even 4-m class telescopes to relatively
bright objects (say $V<12$). Higher dispersions and fainter objects
necessitate even larger apertures, and fortunately spectropolarimetric
instrumentation is present or planned on most of the world's 8-m
facilities.

The origin of linear polarization in a stellar context is usually
dichroic absorption by dust, or scattering by dust, molecules or
electrons (although linearly polarized thermal emission from aligned
grains is sometimes observed in the millimetre regime). For an
unresolved source an asymmetric geometry with respect to the observer's
sightline is required, for example the alignment of grains by a
magnetic field, an asymmetric circumstellar distribution of material
such as a disc, or intra-binary scattering. Cancellation between
polarized radiation from different regions often reduces the observed
polarization to the level of a few percent -- nonetheless the
polarimetry enables one to probe geometrical parameters of an
unresolved object.

In this paper I present a synopsis of some of the stellar
spectropolarimetric studies that my collaborators and I have
undertaken. I concentrate on three broad areas:
Raman-scattering in symbiotic binaries, and its use in measuring
orbits in these long-period systems; the use of recombination-line
polarimetry to measure global structure in the winds of massive stars;
and the use of spectropolarimetric in examining the circumstellar
structure of pre-main-sequence stars. In the final section I give a
brief account of the instruments and likely avenues of research in the
8-m telescope age.

\begin{figure}
\caption{Spectropolarimetry of the 6825\AA\ Raman-scattered line of
  D-type symbiotic RR~Tel. This `triplot' shows the intensity spectrum
  (bottom panel), the polarization magnitude in percent (middle
  panel), and the position angle of the polarization (top panel).}
\label{fig:rrtel_plot}
\includegraphics[width=7cm]{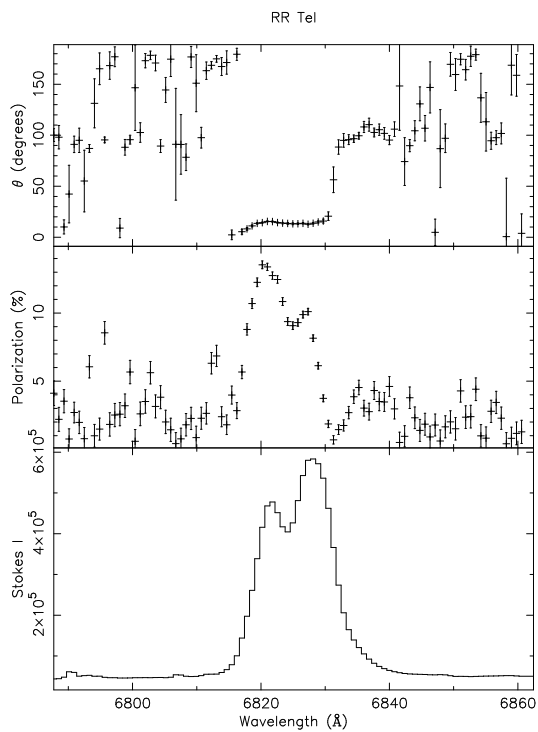}
\end{figure}

\section{Raman scattering in symbiotic binaries}

The spectra of symbiotic stars exhibit both molecular absorption bands
and nebular emission lines. It is now known that this so-called
`combination spectrum' is formed due to binary interaction between a
mass-losing red-giant star and a hot companion (typically a white
dwarf), which ionizes the cool wind from the giant.  Symbiotic stars
are thought to be detached systems \cite{1999A&AS..137..473M}, with
accretion onto the hot component occurring from the wind rather than
via Roche-lobe overflow. Systems that show a stellar spectrum in the
IR are classified as S-type systems, while those that show dust
emission are denoted D-type.
  
In a seminal paper Schmid \cite{1989A&A...211L..31S} determined that
the $\lambda \lambda6825,7082$\AA\ emission lines that are observed in
$\sim$50\% of symbiotic systems arise from Raman scattering of the
$\lambda \lambda1032,1038$\AA\ O\,{\sc vi} resonance doublet by
neutral hydrogen. Raman scattering is the inelastic analogue of
Rayleigh scattering: the frequency of the scattered photon is
different to that of the incident photon, and the scatterer is left in
an altered quantum mechanical state. In this case the O\,{\sc vi}
photons scatter in the far-red wing of Ly$\beta$ and the scattering
H$^0$ atom is left in the $n=2$ state. By conservation of energy the
frequency of the Raman photon, $\nu_r$, is given by
\begin{equation}
\nu_r = \nu_{\rm P} - \nu_{{\rm Ly}\alpha}
\end{equation}
where $\nu_{{\rm Ly}\alpha}$ is the frequency of the Ly$\alpha$
transition and $\nu_{\rm P}$ is the frequency of the parent photon.

\begin{figure}
\caption{The polarimetric orbit of SY~Mus. The O\,{\sc vi} source is
  located at $(0,0)$ and the filled dots show the position of the red
  giant every $0.01P$. Short thick `bowties' indicate the measured PAs
  and their uncertainties, the long thin lines show the calculated PAs
  at the times of the observations. The spectroscopic phases of
  quadrature and conjunction are also marked.}
\label{fig:sy_mus_orbit}
\includegraphics{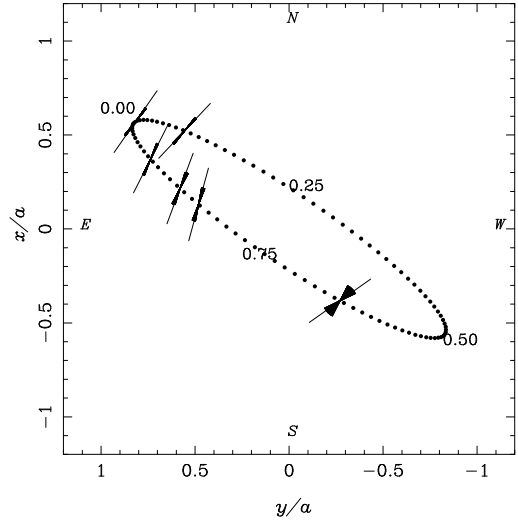}
\end{figure}

The asymmetry of the scattering geometry, with O\,{\sc vi} photons
produced near the hot star scattering in the cool star's wind, coupled
with the dipole nature of the scattering event, leads naturally to
linear polarization \cite{1990A&A...236L..13S}.  Extensive surveys of
symbiotic systems (\cite{1994A&A...281..145S,1996A&AS..119...61H})
revealed that the Raman lines often have a complex structure,
typically comprising two or three peaks in both intensity and polarization
accompanied by a 90$^\circ$ position angle (PA) flip and/or PA
rotation (see, e.g., Figure~\ref{fig:rrtel_plot}).  The symbiotics
show large line polarizations (up to $\sim15$\%), with the strongest
and most highly polarized lines found in D-type (dusty) Mira systems.

Numerical modelling of the Raman-line formation process
(\cite{1996MNRAS.282..511S, 1997A&AS..121...15H,1997MNRAS.292..573L})
indicates that the PA of the blue wing of the Raman-line polarization
is perpendicular to the plane containing the binary line-of-centres
and the observer's line of sight. The models further demonstrate that
the magnitude of the line polarization ($P_l$) is also a simple
function of orbital phase
\begin{equation}
P_l = A \sin^2 \alpha
\end{equation}
where $\alpha$ is the angle between the observer's sightline and the
binary line-of-centres, and $A$ is a scale factor which accounts for
geometrical and physical parameters (interbinary distance, stellar
radii; mass-loss rate, velocity law, etc).  By observing the change in
line polarization with time, it is possible to derive orbital
parameters (e.g., period, inclination, PA of line-of-nodes) with
analogous methods to those used for visual binaries.  The first
`spectropolarimetric orbit', that of SY~Mus
(Figure~\ref{fig:sy_mus_orbit}; \cite{1996A&A...310..235H}), was
followed by those of AG~Dra and Z~And
(\cite{1997A&A...321..791S,1997A&A...327..219S}), while follow-up
observations yielded orbits of four additional systems (CD$-43^\circ
14304$, Hen~1242, M1-21, V455~Sco) \cite{2000A&A...361..139H}, thereby
doubling the number of known orbits with $P>1000$d.

\begin{figure}
\caption{Plots of {\it (a)} orbital period against spectral type, and
  {\it (b)} the distance to the $L_1$ point against stellar radius.
  The targets from this study are plotted as filled squares, and the
  open circles correspond to other symbiotics of `known' period listed
  in \cite{1999A&AS..137..473M}. The dashed line in {\it (a)}
  corresponds to the minimum period relationship while the dotted line
  connects the longest period systems. The dotted lines in panel {\it
    (b)} correspond to $R = L_1$, $R = 2L_1$, and $R = 3L_1$. The
  solid line joins the maximum radius of CD$-43^\circ 14304$
  (100\rsol) to the radius predicted from its spectral type
  (48\rsol).}
\label{fig:symb_period_plot}
\includegraphics[width=\textwidth]{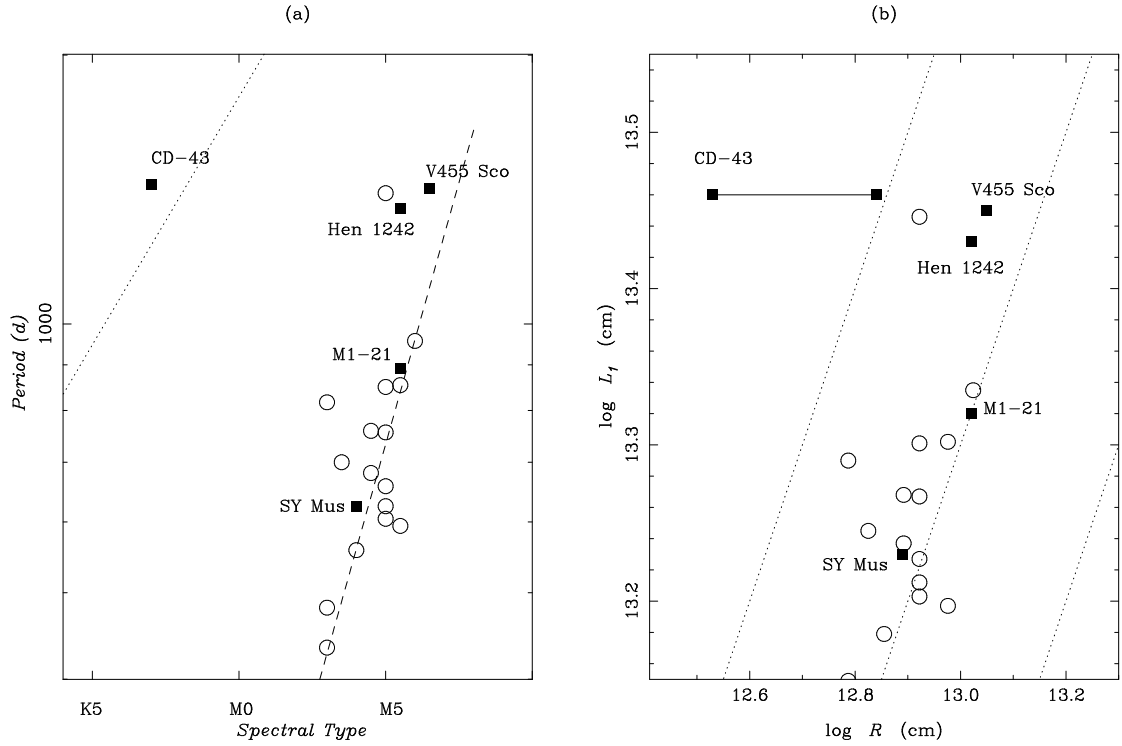}
\end{figure}

It has been shown by M\"urset \& Schmid (\cite{1999A&AS..137..473M})
that there is a strong correlation between orbital period and spectral
type of the cool component, with the majority of systems lying close
to a straight line relating the spectral subtype to the orbital
period. For a given subtype no systems are found with substantially
shorter periods. The minimum period locus is given by
\begin{equation}
\log P_{\rm min} = 0.117S + 2.28
\end{equation}
where $S$ is the M-class subtype and $P_{\rm min}$ is in days (see
Figure~\ref{fig:symb_period_plot}{\it a}).  M\"urset \& Schmid
proposed that this period limit is related to the red giant radius,
which was found to be $L_1/2$ for most systems (where $L_1$ is the
distance from the primary to the inner Lagrangian point). Essentially
this means that for binary to appear as a symbiotic it must be close
enough for significant wind accretion onto the hot component, but wide
enough that the binary is detached
(Figure~\ref{fig:symb_period_plot}{\it b}).  Our systems with
spectropolarimetric orbits conform to this relation, excepting
CD$-43$, which has an exceptionally early spectral-type for its cool
component (K7).
  
The spectropolarimetric approach to orbit determination has unique
advantages.  First, because the observed quantities can be determined
to a small fraction of the orbital amplitudes (e.g., PA to $\sim
1^\circ$ in 180$^\circ$), the approach is extremely efficient -- e.g.,
in the favourable case of SY Mus, we were able to determine correct
orbital characteristics from just four observations, whereas the same
information from traditional radial-velocity techniques required many
observing nights \cite{1994A&A...288..819S}.  Secondly, the amplitude
of variation of the polarization is {\em independent} of
orbital period, so that adequately sampled datasets are capable of
yielding accurate orbital characteristics for systems with periods of
from a few months to many years (unlike normal spectroscopic methods).
Our group, along with Schmid's at ETH Z\"urich, are continuing to
obtain data on various symbiotics with the aim of establishing periods
for many more S-type systems, and obtaining the first reliable D-type
orbit. The latter goal is obviously long-term, since D-type systems
are likely to have periods of many decades!

\section{The geometry of stellar winds}

Massive stars shed mass via high-speed, radiation-driven winds.
Mass-loss rates of $10^{-7}$ to 10$^{-4}$ M$_\odot$\,yr$^{-1}$ are
typical, with terminal speeds of 1000--3000\,km\,s$^{-1}$. Stellar
wind models are often used to fit recombination line profiles (such as
H$\alpha$) in order to determine mass-loss rates. This approach
essentially reduces to `counting' recombinations, based on the
standard assumptions of homogeneity and sphericity. Departures from
either will lead to an overestimate of the true mass-loss rate: a wind
that is clumpy on small scales, or is global distorted, will have
an increased rate of recombinations (which is a
density-squared process) per unit mass, so a structured wind will
produce greater line emission than a smooth wind with the same
mass-loss rate.

\begin{figure}
\caption{Spectropolarimetry of the WN star WR~134 and the WC star
  WR~137, binned to a constant polarization of 0.1\% and 0.05\%
  respectively. Note the strong depolarizations at the emission line
  wavelengths.}
\label{fig:wr_plot}
\includegraphics{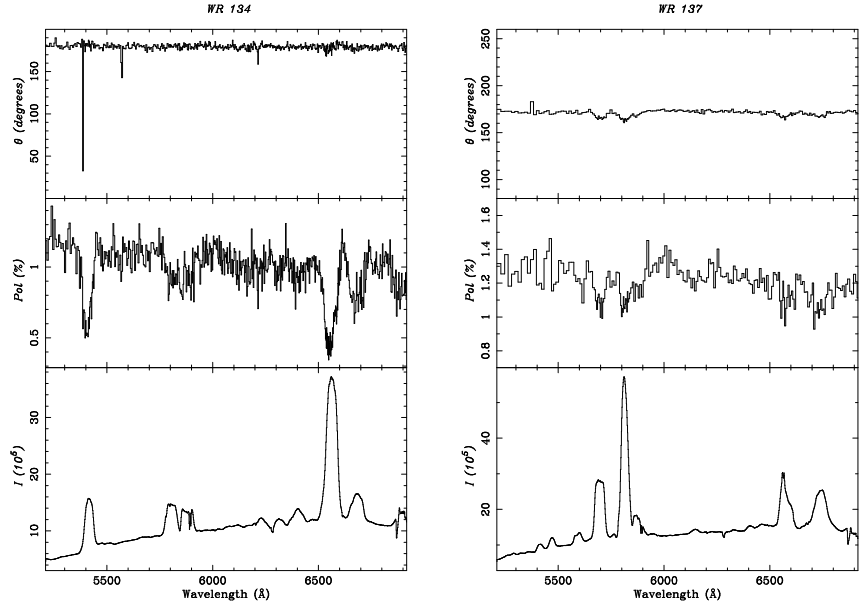}
\end{figure}

So how applicable is the assumption of sphericity? The winds cannot be
imaged directly, but fortunately polarimetry provides an indirect
probe of the geometry of the wind near to the star. Continuum photons
from the star will pass through the ionized wind, and some will be
electron-scattered which introduces linear polarization to the
starlight. The nett polarization will be zero for a spherical
distribution of electrons, but will be non-zero for any geometry that
is not circularly symmetric with respect to the observer's line of
sight.  The recombination line emission, which is produced further out
in the wind, will `see' a lower electron-scattering column, and will
therefore be less polarized than the continuum. Spectropolarimetry of
a non-spherical wind should therefore show a polarized continuum (on
the order of a few tenths of a percent) with the magnitude of the
polarization dropping at the emission line wavelengths, as the
polarized continuum is diluted by unpolarized line emission. This
known is the `line effect'. As an illustration I have plotted
polarization spectra of two Wolf-Rayet (W-R) stars in
Figure~\ref{fig:wr_plot}: note the obvious decline in continuum
polarization magnitude across the emission lines, a sure indication
that the stellar winds in these objects depart from spherical
symmetry.

\begin{figure}
\caption{The 1998 $K'$ maximum entropy restored images of WR137
  (logarithmic greyscale), and the 1997 image (contours) adapted from
  Fig. 3 of \cite{1999ApJ...522..433M}. The dashed line illustrates
  the PA of the intrinsic polarization.}
\label{fig:kband_fig}
\includegraphics{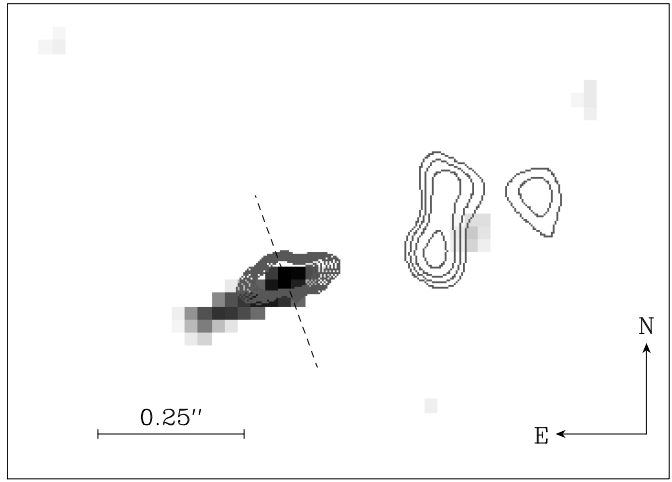}
\end{figure}

For WR137 we have other indicators that wind may be asymmetric. This
WC star has an O-star binary companion in an eccentric orbit with a
period of $\sim 12$ years. At periastron the interaction of the winds
from the two components provides the necessary conditions for the
condensation of warm carbonaceous dust grains, which produce a near-IR
excess \cite{2001MNRAS.324..156W}. There is an intimate relationship
between the solar-radius-scale geometry of the stellar wind from
polarization, and the larger scale distribution of newly formed dust
(many hundreds of AU) -- see Figure~\ref{fig:kband_fig} and
\cite{2000A&A...361..273H}.  The link between near-star geometry and a
large-scale nebula has also been established for the Luminous Blue
Variable (LBV) AG~Car \cite{1994ApJ...429..846S}. Spectropolarimetry
of this object shows a strong line effect at H$\alpha$, along with
substantial variation ($>1$\%) in continuum polarization. The PA of
the intrinsic polarization is aligned with the major axis of the
AG~Car ring nebulae and perpendicular to its jet, indicating that the
resolved circumstellar environment is already present within a few
stellar radii of the central object.

The magnitude of the continuum polarization ($P_c$) for a given
stellar wind geometry and viewing angle ($i$) may, in the optically
thin limit, be computed analytically \cite{1977A&A....57..141B}:
\begin{equation}
P_c \approx 2 \overline{\tau} (1 - 3\gamma) \sin^2 i
\end{equation}
where $\overline{\tau}$ is the angle-averaged electron-scattering optical depth
of the envelope and $\gamma$ is a shape factor. The degeneracy between
the stellar wind geometry and the viewing angle means that studying
objects on an individual basis is rather difficult, but if a large
enough sample is obtained it is possible to disentangle the two
statistically. We gathered a spectropolarimetric data set of 16~W-R
stars using the 4-m William Herschel Telescope (WHT) and the ISIS
spectrograph \cite{1998MNRAS.296.1072H}. A statistical analysis of these
data using Monte-Carlo based K-S tests enabled us to show that about
15\% of W-R stars show a global departure from spherical symmetry. The
level of the intrinsic polarization is consistent with an
equator-to-pole density ratio of 2--3. We postulated that the
line-effect systems represented the most rapidly rotating W-R stars, in
which the rotation was having a significant affect on the wind
dynamics.

\begin{figure}
\caption{Wolf-Rayet stars plotted in the mass-loss/luminosity
  plane. WN stars are plotted as squares, while WC stars are plotted
  as triangles. Stars with a line effect are displayed as open
  symbols, stars with no detectable line effect are shown as filled
  symbols.}
\label{fig:lmdot_plot}
\includegraphics{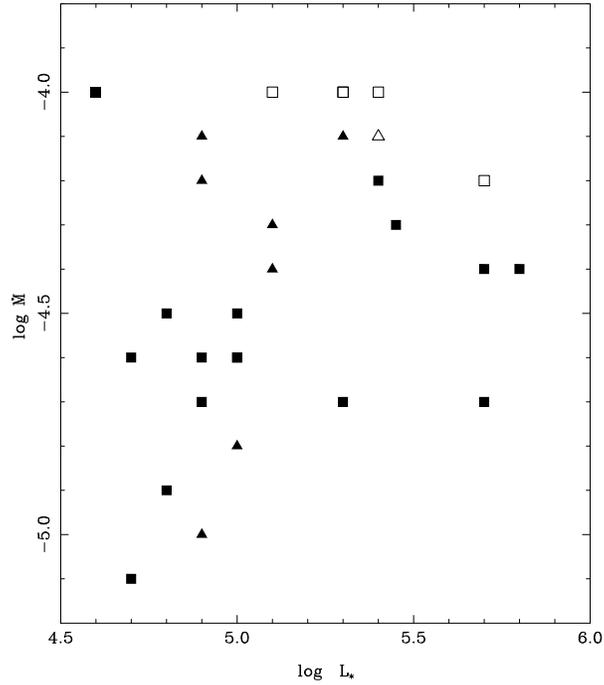}
\end{figure}

Interestingly the line-effect systems are clustered towards high
mass-loss rates and high-luminosity (Figure~\ref{fig:lmdot_plot}).
Although at first sight this appears to suggest that this may be the
root cause for the wind asymmetry, one should note that these
quantities are very dependent on small and large-scale structure, as
noted above. In other words the high mass-loss rates and luminosities
determined by the models may be a result of applying spherically
symmetric models. Mass-loss rates derived from thermal radio emission,
which are the least model-dependent, show good agreement with those
derived from the optical recombination lines.  This does not mean that
the mass-loss rate values are correct in an absolute sense, since both
the radio continuum and the recombination lines scale as
density-squared, but the agreement does indicate that the degree of
wind asymmetry is similar at the radii for optical depth $\sim 1$ of
both the optical $\sim 10^1 R_*$ and the radio $\sim 10^3 R_*$.

Since departures from spherical symmetry appear to be relatively
common for the W-R stars, and also occur in LBVs, it seemed natural to
search for evidence for global structure in the winds of
O-supergiants.  We obtained H$\alpha$ spectropolarimetry of a sample
of 20 Northern, and combined these data with $K$-band
spectropolarimetry obtained with the 4-m United Kingdom Infrared
Telescopes and low resolution optical spectropolarimetry from Pine
Bluff Observatory \cite{2002MNRAS.337..341H}. Somewhat surprisingly,
the spectropolarimetry revealed that the single O-supergiants
typically have intrinsic polarizations of $< 0.1$\% (which
corresponds to an equator:pole asymmetry of approximately less than
1.25). It appears that the wind line-driving mechanism in
rapidly-rotating O~supergiants (whose winds are optically thin) is
relatively insensitive to rotation compared to the optically-thick
line-driving mechanism of W-R outflows.

W-R stars are also known to have small-scale clumps that appear to
accelerate slowly from the star (e.g. \cite{1996ApJ...466..392L}), and
there is spectroscopic evidence for similar structures in the
O~supergiant $\zeta$~Puppis \cite{1998ApJ...494..799E}. These
migrating density structures could potentially be identified
polarimetrically, but numerical models of density enhancements
propagating through the wind predict polarizations of close to
0.1\%--0.2\% \cite{2000MNRAS.315..722H}, which is uncomfortably close
to the current limits on the absolute precision of polarimetric
observations.  It is clear that very precise techniques will be needed
to probe these clumps in the winds of O~supergiants.

\section{Pre-main-sequence stars}

The broad picture of low-mass star formation, from the collapse of a
gravitationally unstable molecular cloud, through to a
pre-main-sequence star surrounded by a dusty accretion disc, is
well-established. During the final, classical T~Tauri star (CTTS)
phase magnetospheric accretion is thought to occur, with the star's
magnetic field disrupting the disc and channelling the material onto
the surface, where its kinetic energy is liberated at hot spots.
However, the situation for intermediate and high-mass star formation
is much more confused, with a lack of clear-cut evidence for
circumstellar discs. In the previous sections I have shown that line
spectropolarimetry can be a powerful probe of the circumstellar
geometry, and we are now in the process of applying
spectropolarimetric techniques to pre-main-sequence (PMS) stars.

We have obtained a large sample of H$\alpha$ polarization spectra
Herbig Ae/Be stars \cite{2002MNRAS.337..356V}. The two classes of star
show a dichotomy in spectropolarimetric characteristics. The Herbig Be
stars typically show the classical depolarization line effect, which
is indicative of a large volume of H$\alpha$ emitting material
embedded in an asymmetric envelope. The Herbig Ae stars on the other
hand show a more complex polarization through the line, with some
stars showing significant polarized H$\alpha$ emission. This suggests
that Herbig Ae stars have compact sources of H$\alpha$ emission,
possibly magnetospheric accretion streams, with H$\alpha$ photons
scattering in the surrounding asymmetric medium. It is not completely
clear whether the scatterers here are dust or electrons, although
thermal motions of electrons should lead to significant smearing of
the line polarimetry, which is not observed. Our results indicate that
the magnetospheric accretion paradigm for low-mass star formation may
extend to higher masses.

\begin{figure}
\caption{H$\alpha$ spectropolarimetry of a Herbig Be star (HD~53367) and a Herbig
Ae (XY Per), binned to a constant error in polarization of 0.03\% and
0.1\% respectively. Note that the Herbig Be star shows a classical
line depolarization, whereas the polarization spectrum of the Herbig
Ae star is more complex.}
\label{fig:herbig_plot}
 \includegraphics[width=\textwidth]{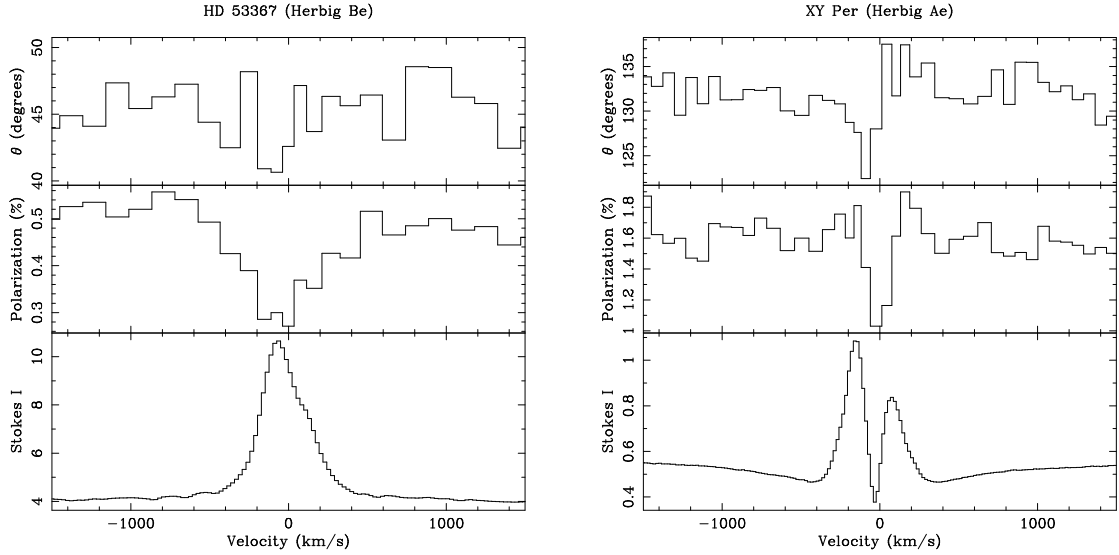}
\end{figure}

\begin{figure}
\caption{H$\alpha$ spectropolarimetry of the classical T~Tauri star
  RY~Tau, binned to a constant error in polarization of 0.1\%.}
\label{fig:ttauri_plot}
 \includegraphics{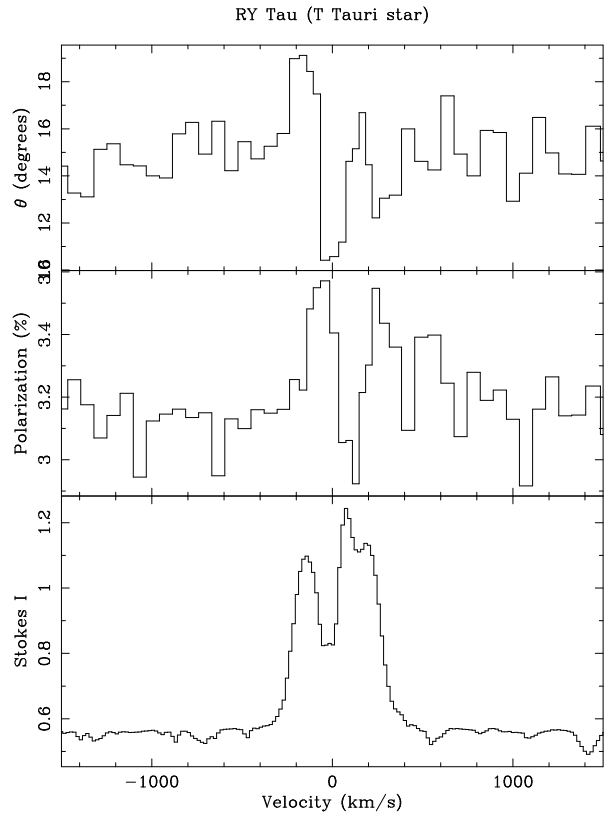}
\end{figure}

A significant investment has been made in photopolarimetric studies of
classical T~Tauri stars (CTTS), which may show continuum polarizations
of a few percent e.g.
\cite{1982A&AS...48..153B,1985ApJS...59..277B,1992AJ....103..564M,
  1996A&A...308..821G}. This polarization is variable, perhaps due to
changes in the illumination of the scattering envelope due to hot
spots \cite{1998ApJ...506L..43W}. Since it was thought that the line
and continuum photons from the central object were scattered by the
same dusty envelope the expectation was that there should be no line
effect. However our recent observations of RY~Tau
(Figure~\ref{fig:ttauri_plot}, \cite{2003A&A...406..703V}) show a
dramatic change in the polarization through H$\alpha$. It appears that
RY~Tau is not exceptional in this regard: in our recent survey 9 out
of 10 CTTS showed a detectable change in polarization across H$\alpha$
\cite{vink2005b}. These observations are qualitatively consistent with
scattering of a compact source of H$\alpha$ photons off a
circumstellar disc.

A more detailed interpretation of the spectropolarimetry requires
numerical modelling. As a first attempt we have examined the
polarimetric signature of dipolar scattering of a compact line source
off optically-thin and optically-thick dusty discs \cite{vink2005}
using the Monte-Carlo radiative-transfer code {\sc torus}
\cite{2000MNRAS.315..722H}. We find that the PA dependence of the line
profile is an indicator on the extent of the disc, with a double PA
rotation through the line corresponding to a disc with a negligible
inner hole, and a single PA rotation indicating a substantial
evacuated area around the central star. By interpreting the
observations in the context of these numerical simulations we will
begin to place strong constraints on the circumstellar environment of
PMS across the mass spectrum.

\section{The future}

Spectropolarimetric capabilities are already available on 8-m class
telescopes.  For example, VLT/FORS offers both circular and linear
spectropolarimetry, but at $R\sim 2500$ its maximum resolution is
insufficient for most of the line polarization work described here
(although the broad emission lines in supernovae can be resolved).
Similarly the Keck/LRIS combination is limited essentially to
continuum studies and is used predominantly for extragalactic
research. The two 8-m Gemini telescopes are due to be fitted with
facility polarization modulator units (GPOL), which will eventually
provide a number of their optical and near-IR instruments, including a
high-dispersion echelle spectrograph, with a polarimetric capability.
The Prime Focus Imaging Spectrograph (PFIS) on the 10-m South African
Large Telescope will provide a medium resolution ($R \sim 5000$)
spectropolarimetric capability, and is ideally suited to stellar work.

Pleasingly it appears that spectropolarimetry is now viewed as a
standard mode by spectrograph developers. The proliferation of
spectropolarimetry in the 8-m telescope age will lead to exciting new
advances in stellar astrophysics, for example in the investigation of
the link between rotation and outflow asymmetries and structure in
massive stars and asymmetries in core collapse supernovae, which can
show a line effect (\cite{1988MNRAS.231..695C,2003ApJ...592..457W}),
and may have viewing-angle-dependent observed luminosities. Rotation
and asymmetry obviously play a key role in the collapsar model for
gamma ray burst progenitors \cite{1999ApJ...524..262M}, and
high-resolution linear spectropolarimetry of the burst afterglows
should yield valuable insights. Extending the spectropolarimetric
study of W-R stars into the low-metallicity environment of the
Magellanic clouds would provide a good test of line-driving theories.
The move to longer wavelengths will also lead to new discoveries, for
example by probing the circumstellar environments of highly obscured
objects, such as Class~I protostars that are still embedded in their
nascent molecular clouds.


\begin{theacknowledgments}
  I am grateful to the SOC for the invitation and for hosting a
  pleasant and interesting meeting. I am of course indebted to my
  research collaborators in polarimetry: Ian Howarth (UCL, UK), Ron
  Hilditch (St.Andrews, UK) John Hillier (Pittsburgh, USA), Regina
  Schulte-Ladbeck (Pittsburgh, USA), Janet Drew (Imperial College,
  UK), Jorick Vink (Imperial College, UK), Ren\'e Oudmaijer (Leeds,
  UK), Ryuichi Kurosawa (Exeter, UK), Neil Symington (Exeter, UK),
  Bruce Babler (Wisconsin, USA), and J-F Donati (Toulouse, France).
\end{theacknowledgments}


\bibliographystyle{aipproc}   

\bibliography{harries}

\IfFileExists{\jobname.bbl}{}
 {\typeout{}
  \typeout{******************************************}
  \typeout{** Please run "bibtex \jobname" to optain}
  \typeout{** the bibliography and then re-run LaTeX}
  \typeout{** twice to fix the references!}
  \typeout{******************************************}
  \typeout{}
 }

\end{document}